\documentclass[12pt]{article}

\usepackage{epsfig}
\usepackage{amsmath,amsfonts,amssymb}

\newcommand{\unit}{\leavevmode\hbox{\small1\kern-3.6pt\normalsize1}}

\def\lsim{\raise0.3ex\hbox{$\;<$\kern-0.75em\raise-1.1ex\hbox{$\sim\;$}}}
\def\gsim{\raise0.3ex\hbox{$\;>$\kern-0.75em\raise-1.1ex\hbox{$\sim\;$}}}

\def\asusy{a^{\rm SUSY}_\mu}

\def\bmumu{B_s^0\to\mu^+\mu^-}

\newcommand{\captions}{\sf\caption}

\allowdisplaybreaks

\begin{document}

\thispagestyle{empty}
\begin{flushright}
  FTUAM 09/16\\
  IFT-UAM/CSIC-09-32\\
  CPHT-RR077.0709 \\
  LPT-Orsay 09/59\\

  \vspace*{2.mm}{29 July 2009}
\end{flushright}

\begin{center}

{\Large {\bf Gravitino dark matter in hybrid gauge-gravity models}} 
\vspace{1 cm}\\

{\large D.G.~Cerde\~no $^{1}$, Y.~Mambrini $^2$,  A.~Romagnoni $^{2,3}$}
\vspace{1cm}\\

{\textit{
    $^1$
    Departamento de F\'{\i}sica Te\'{o}rica C-XI,
    and Instituto de F\'{\i}sica Te\'{o}rica UAM-CSIC, \\[0pt]
    Universidad Aut\'{o}noma de Madrid, Cantoblanco, E-28049 Madrid,
    Spain
    \vspace{0.3cm}\\
    $^2$
    Laboratoire de Physique Th\'eorique,
    Universit\'e Paris-Sud, F-91405 Orsay, France
    \vspace{0.3cm}\\
    $^3$
    Centre de Physique Th\'eorique, Ecole Polytechnique, CNRS,
91128 Palaiseau, France
    \vspace{0.3cm}\\
}}

\end{center}

\vspace{0.7cm}

\begin{abstract}
  We study the phenomenology of generic supergravity
  models in which gravity
  mediation naturally competes with gauge mediation
  as the origin of supergravity-breaking.
  This class of hybrid models has been recently motivated in string
  inspired constructions
  and differs from usual gauge mediated supersymmetry breaking 
  models in having messenger masses
  of order of the GUT scale. In these scenarios the gravitino can be
  the lightest supersymmetric particle in wide regions of the
  parameter space and therefore a potential 
  candidate for dark matter. 
  We investigate this possibility, imposing the WMAP bound on its
  relic 
  abundance and taking into account constraints from Big Bang
  nucleosynthesis.
  We show that in these constructions viable gravitino dark matter can
  be obtained
  in specific regions of the parameter space,
  featuring large values of
  $\tan \beta$ and where the supersymmetry breaking mechanism is
  dominated by gauge mediation.  
\end{abstract}

\newpage

\section{Introduction}

In the largest class of viable scenarios of supersymmetry (SUSY)
breaking
a hidden sector is present where the breaking of SUSY takes place,
which
is then communicated to the 
visible
one by mediator fields 
via loop-suppressed or nonrenormalizable interactions. It is possible
to parameterize the hidden sector 
dependence on the breaking by $\langle F_{\phi_i} \rangle$, the vacuum
expectation value ($vev$)
of the auxiliary component of the spurions fields $\phi_i$. 
The resulting mass scale of 
scalars and gauginos in the visible sector is related to the
messenger scale with contributions of the order 
$\sum_{i} a_i \langle F_{\phi_i} \rangle/ \langle \phi_i
\rangle$, where $a_i$ are model dependent parameters which can
generically be loop suppressed. 
Depending on whether the messenger fields which transmit the 
SUSY breaking
into the observable sector have only gravitational or also gauge
interactions, the mechanism is described as
gravity mediated or gauge mediated SUSY breaking (GMSB)
respectively.

In this work we concentrate on scenarios which mix both
kinds of mediations, 
called hybrid models. Their formulation, already considered from a
model building perspective in Ref.\,\cite{Poppitz}, has been recently
motivated also as effective field models of string theory
constructions in Ref.\,\cite{Dudas:2008qf}. In these cases we can
distinguish two classes of spurion fields responsible for the breaking
of SUSY in the hidden sector: moduli fields, $T_i$, interacting
with the Standard Model (SM) sector through 
gravitational interactions, and
singlet chiral superfields, $X_j$, which couple directly to messenger
fields in the superpotential. We can sum up these interactions
and for simplicity consider for our purposes the following
superpotential, depending on just one modulus $T$ and one spurion
field $X$
\begin{equation}
  W = \lambda X M \overline M +W(X,T) \ .
  \label{W}
\end{equation}
Here $M$ and $\overline{M}$ are messenger superfields
with $SU(3)\times SU(2) \times U(1)$ quantum numbers, whereas $T$ and
$X$ are SM-gauge singlets. In this framework the supersymmetric mass
for the messenger fields is clearly proportional to $\langle X
\rangle$.  We are considering a superpotential with just
renormalizable (and then minimal) couplings with the spurion field
but obviously it can be further generalized, as well as the
K\"ahler potential, for which we adopt here the standard expression.

Since the messengers lie within a representation of the SM
gauge interactions,  
gaugino masses and scalar squared-mass parameters appear at one
and two loops, respectively, 
being the resulting supersymmetric masses 
of the order \cite{Giudice:1998bp} 
\begin{equation}
  M_{\mathrm{GMSB}}^{i} \sim \left( \frac{a_i}{16 \pi^2} \right)
  \left( \frac{F_X}{X} \right), 
  \label{Msusy}
\end{equation}
where $i$ denotes  
the gauge index representation for the scalar or gauginos
and $a_i$ are coefficients of order unity. Concerning the
gravitino, its mass is related to the fundamental SUSY breaking mass
by the null cosmological constant condition $\langle V \rangle = 0$,
\begin{equation}
\sqrt{3} m_{3/2} M_{\mathrm{Pl}}
=\sqrt{|F_{T}|^2 + |F_X|^2}.
\label{m3/2}
\end{equation}

Generically, GMSB models lead to small values for the gravitino mass
if the  only source of SUSY breaking comes from the spurion field. 
Indeed, if the dominant term of Eq.(\ref{m3/2}) is $|F_X|$, using
Eq.(\ref{Msusy}) we see that in order to have a MSSM mass
spectrum of order of the TeV, the
gravitino mass reads
\begin{equation}
  m_{3/2} \sim \frac{16 \pi^2 \langle X \rangle M_{\mathrm{GMSB}}^{i}
  } {M_{\mathrm{Pl}}} \sim 10^{-13} \langle X \rangle\ ,
  \label{m3/2bis}
\end{equation}
with  $M_{\mathrm{Pl}}=(8 \pi G_N)^{\frac{1}{2}}\sim 2.4 \times
10^{18}$ GeV and  $\langle X \rangle \sim M_{\mathrm{Mess}}$. 
We see that in the usual GMSB case, where the messenger masses are
of order of 100\,TeV, the resulting gravitino mass
is of the order of 1\,eV. Heavier gravitinos are allowed if one
assumes other dominant sources 
for the SUSY breaking which generates the gravitino
mass (for example $|F_T|$ in models with secluded
sectors). This adds new degrees of freedom and alters the direct 
proportionality of Eq.\,(\ref{m3/2bis}) between the gravitino mass and
$\langle X\rangle$.

Recently another possibility was proposed in
Ref.\,\cite{Dudas:2008qf}, namely 
that the dynamics of the model forces the spurion $X$ to be
stabilized at a near-GUT scale.  
In fact in this type of models, the presence of a Fayet-Illiopoulos
term, $\xi$, generated at a string scale (typically $\sim 10^{-1}-
10^{-3} M_P$), requires that $X$ takes a $vev$ which cancels the
$\xi$ contribution to the potential, through the D-term flatness
condition. 
In other words, the dynamics of the theory pushes $\langle X\rangle$
towards large values of the order of $10^{16}$~GeV. However, in order
to obtain a MSSM spectrum within the TeV range, a 
fundamental scale $\Lambda = F_X/X \sim 10^5$ GeV is needed. 
This in turn requires higher  
values for $F_X$, and consequently heavier gravitinos, $m_{3/2} \sim
10^3$~GeV, also implying the
interference of gravity mediation with the gauge mediation
mechanism. This class of 
constructions was called hybrid models, 
which were introduced in Ref.\,\cite{Poppitz}.
The phenomenological consequences
are numerous and
``interpolate" between gravity-mediated models (such as the usual
Constrained MSSM) and GMSB scenarios.

In this sense, a very 
appealing feature of supersymmetric theories is that they can
provide candidates to solve the problem of the dark matter in the
Universe in terms of the
lightest supersymmetric
particle (LSP). A discrete symmetry, $R$-parity, is
often imposed in order to forbid lepton and baryon violating
processes which could lead, e.g., to rapid proton
decay. This implies that  SUSY particles
are only produced or destroyed in pairs, thus rendering the LSP
stable.
Among the most interesting possibilities for supersymmetric
dark matter are the lightest neutralino
\cite{Goldberg:1983nd,Munoz:2003gx}, which enters the category of
weakly-interacting massive particle,
and the gravitino \cite{gravitino}, 
which only has gravitational couplings and is
therefore extremely weakly-interacting.

The viability of gravitino dark matter has been widely
studied within the
context of supergravity models in which the gravitino mass enters as a
free parameter. These supergravity scenarios can be thought of as
appearing as the low-energy limit of some more fundamental string
models. However, in most of the stringy inspired scenarios
studied so far the gravitino is not the LSP\footnote{
  Notice in this sense that 
  a class of 6D chiral gauged supergravity was studied in 
  \cite{Choi:2009jd} which presents regions with gravitino LSP.}.
As we will show,
this is not the case in hybrid models, in which both the neutralino
and gravitino can be the LSP in different areas of the parameter
space.
The regions with neutralino dark matter were already studied in
Ref.\,\cite{Dudas:2008qf} but the possibility of gravitino dark matter
has not been addressed yet.
In this work we investigate the viability of the gravitino as a dark
matter candidate in this class of hybrid models, calculating its relic
abundance and imposing Big Bang nucleosynthesis (BBN)
constraints.

The paper is organized as follows. In Section\,2
we summarize the parametrization used in our phenomenological analysis
and we discuss the theoretical motivations and the peculiarities of the
hybrid models with respect to the usual gravity and gauge mediated
scenarios. In Section\,3 we explore the conditions under which the
gravitino can be the LSP and a good dark matter candidate. Finally, in
Section\,4 we expose our conclusions.

\section{The model}

The contribution from the two different mediation mechanisms we are
considering can be parametrized in a general way
by the gravitino mass, $m_{3/2}$, and two dimensionless
parameters, $\alpha$ and $\delta$, 
which measure the relative sizes of
standard (F-term induced) and non-standard (D-term induced
\cite{Poppitz:1996xw}) gauge mediation contributions
in units of $m_{3/2}$ respectively. The latter, 
in particular, have to be taken into account when 
extra abelian gauge groups enter in the computation 
(see, e.g., Ref.\,\cite{Dudas:2008qf} for details).
The soft supersymmetry-breaking terms can then
be written as
\begin{eqnarray}
  M_a &=& M_a^{\mathrm{Grav}} + M_a^{\mathrm{GMSB}}=
  m_{3/2} \left( \tilde{\epsilon} + g^2_a S_Q~ \alpha \right)\,,
  \nonumber \\
  m_i^2 &=& (m_i^{\mathrm{Grav}})^2+ (m_i^{\mathrm{GMSB}})^2=
  m_{3/2}^2 \left(1+ C_i S_Q  ( \delta + \frac{\alpha^2}{N}) \right)\,,
  \label{general}
\end{eqnarray}
where $N$ is the
effective number of messenger fields 
contributing to gauge mediation,
$S_Q$ is the Dynkin index of the messenger representation
($1/2$ for the fundamental representation of $SU({\cal N})$),
$g_a$ are the gauge couplings and
$C_i = \sum_{a} g^4_a C^a_i$,  $C^a_i$
being the Casimir of the MSSM scalar fields representations
(in our normalization the Casimir of the fundamental representation of
$SU({\cal N})$ is $({\cal N}^2-1)/(2{\cal N})$,  that of  $U_Y(1)$ is
simply $Y^2$). 
In our phenomenological analysis we consider a
flavor universal case, where the gravity-mediated contributions
are dominated by the term
\begin{equation} \label{fl-univers}
(m_{ij}^{\mathrm{Grav}})^2  \simeq
m_{3/2}^2\, \delta_{ij}\,,
\end{equation}
keeping in mind that in principle 
there could be some flavor-mixing effects from the gravity side.
However this assumption is justified in generic supergravity 
constructions.
The extra parameter $\tilde{\epsilon}$ includes the effects
of gravity mediation for gauginos. In this case the gravitational
contributions are present only if the gauge kinetic function depend
on the modulus field $T$. Moreover, since these contributions are
proportional to the ratio $F^T/T$, in the cases under consideration in
our analysis
this universal coefficient is naturally of order
$\tilde\epsilon\sim{\cal O} (10^{-1})$. It is therefore 
suppressed with respect to the above mentioned universal
contribution from gravity mediation to the scalar masses (which is of
order 1). 
Different values
should be taken into account in the cases where extra (for instance
secluded) sectors are included in the model.

Unlike the classical GMSB at low energy, gauge mediation in hybrid
models occurs around the GUT scale, where the gauge contributions to
the gaugino masses $M_a$ (proportional to their gauge couplings $g_a$)
are approximately gauge universal.
Thus, the gauge non-universality only affects 
scalars masses. Concerning the trilinear couplings $A_{i=t,b,\tau}$,
there is no 1-loop messenger contribution to the SUSY-breaking
trilinear terms. However, $A_i$ terms are generated in the leading-log
approximation by the RG evolution and are proportional to gaugino masses.
For simplicity, in our analysis we will assume that 
the trilinear terms are universal at
the GUT scale and given by a unique parameter $A$. In order to study
the effects of variations in the trilinear term, 
we will consider the two examples with
$A=0$ and $A=-3 m_{3/2}$.
The reader can find in the appendix of Ref.\,\cite{Dudas:2008qf} the
explicit expressions
of the mass terms for each generation of squarks and sleptons.

Such a hybrid model has several peculiarities.
\begin{itemize}
\item{The value of the 
  GUT scale for the messengers sector appears
  in a natural way. In fact, the dynamics of supersymmetry breaking
  itself justifies  very heavy masses for the messengers (of the order
  of the Fayet-Illiopoulos term for instance) and the rest of the
  spectrum, at least qualitatively, turns out to be strongly
  constrained by this first peculiarity. Moreover, as it was pointed out 
  for example in Ref.\,\cite{Dudas:2008qf}, it seems difficult to avoid
  hybrid scenarios in any stringy inspired supergravity scenario with
  extra $U(1)$.}

\item{The regions with viable neutralino dark matter and allowed by
  WMAP constraints have quite distinctive phenomenological
  consequences which in principle could be observable at LHC
  \cite{Dudas:2008qf}. For example, the measurable
  non-universality in the scalar soft breaking terms makes it possible
  to distinguish this scenario from the Constrained MSSM and the fact
  of generating trilinear couplings makes it possible to distinguish
  it from pure GMSB.}

\item{The FCNC problem, inherent to gravity mediated supergravity
  constructions, is 
  alleviated by the gauge mediated contributions. In particular, for
  large values of $\alpha$
  the assumption made in 
  Eq.\,(\ref{fl-univers}) concerning the flavor dependence of the
  gravity contribution is not so relevant since the gauge mediated
  contribution dominates. Interestingly, as we will see in the next
  Section, it is precisely in this region of the parameter space
  that the gravitino is a viable dark matter candidate.}

\item{The gravitino is naturally heavy (TeV scale) without the need of
  extra supersymmetry breaking sectors. Indeed,
  from Eq.(\ref{m3/2bis}) we clearly see that heavy messengers
  (i.e., large values of
   $\langle X \rangle \sim M_{GUT}$) can easily be consistent with
  large values for $m_{3/2}$.} 

\item{Concerning the $\mu/B\mu$ problem characteristic of pure GMSB
  constructions, hybrid models can help finding a solution through the
  Giudice-Masiero mechanism  \cite{Giudice-Masiero}. One of the main
  problems arising in gauge mediation  constructions, where the Higgs
  fields directly couple to the spurion, is the fact that it is
  difficult to satisfy the MSSM-induced relation 
  \begin{equation}
    \sin 2 \beta = \frac{B \mu}{m_{H_1}^2+m_{H_2}^2+2 \mu^2} \sim
    \frac{\mu \Lambda}{m_{H_1}^2+m_{H_2}^2+2 \mu^2},
    \label{sin2beta}
  \end{equation}
  (where $m_{H_i}$ are the Higgs soft masses) due to the fact that  
  $\Lambda \gg
  \mu$. However, hybrid models can address
  this issue, giving an extra gravitational contribution to the $\mu-$term. 
  Indeed, one can show
  that if a K\"ahler term
  of the form
  \begin{equation}
    K=\int d^4 \theta ~
    Z(T, \bar{T}, X, \bar{X}) H_1 H_2,
  \end{equation}
  is introduced,
  where $Z(T, \bar{T}, X, \bar{X})$ is a modular function ensuring the
  modular invariance of the term $Z(T, \bar{T}, X, \bar{X}) H_1 H_2$,
  one can generate a $\mu$
  and $B\mu-$term after SUSY breaking of the order
  \begin{equation}
    \mu \sim  m_{3/2} \langle Z  \rangle\ ,  \nonumber \qquad \qquad  B
    \mu \sim 2 ~m_{3/2}^2  \langle Z \rangle\ . 
  \end{equation}
  Then the values of $m_{H_1}$ and $m_{H_2}$ (including their
  contributions from GMSB) can be easily arranged to fulfil 
  $\sin 2\beta <
  1$. Notice that in this case a direct coupling between the Higgs
  and the spurion fields $\lambda' X H_1 H_2$, would require an
  unreasonably small 
  coupling $\lambda' \sim 10^{-16}$  to obtain a TeV
  scale $\mu-$term. However, 
  such a direct coupling can be easily avoided imposing, for example,
  suitable charges for the Higgs fields under an extra abelian gauge
  group. Even in the absence of such an interaction, the supergravity
  sector provides a $\mu-$term and $B\mu-$term of the right order of
  magnitude since the gravitino mass is already approximately 
  $100$ GeV to $ 1 $TeV.

  Notice however that
  large values of $\alpha$ imply the dominance of gauge mediation
  contributions, which implies that
  gauginos and  
  scalars are at the same mass scale. In particular, we will show in
  the next Section that for a  
  gravitino of order $100$~GeV and $\alpha \sim 50 - 100$, 
  the whole soft
  spectrum is very heavy, and  
  the resulting values for $\mu$ are of order of the TeV.  
  This implies a tension, from a theoretical point of view, with the
  predicted value of $\mu$ from the  
  Giudice-Masiero mechanism. 
  A possible way-out could be
  provided by the presence of  
  the non-standard gauge mediation contributions parametrized by
  $\delta$. In fact, $\delta$  
  acts just for the scalar mass contributions and could then help in
  approaching the so-called  
  ``focus-point" region where a small $\mu$ is expected. In any case, 
  it is
  obvious that this kind of  
  solution needs a $\delta$ at least comparable with
  $\frac{\alpha^2}{N}$, which seems unnatural  
  in the class of UV model considered here as
  examples. 
}

\item{ The contributions from anomaly mediation \cite{anomaly} can be
  neglected in this kind of models. In fact, since these are
  proportional to $\frac{g^2}{16 \pi^2} m_{3/2}$, they 
  are naturally loop-suppressed with respect to the universal terms
  generated by 
  the gravity mediation, for $\tilde{\epsilon} \sim {\cal
    O}(10^{-1})$. Therefore we will not consider effects like mirage
  mediation \cite{Lebedev}, 
  or deflected mirage mediation \cite{deflected}, 
  which appear when the anomaly and gravity contributions are of the  
  same order because of the suppression of the gravity
  mediation due to the moduli 
  couplings. 
  Among other things, from a phenomenological point of
  view this means that in hybrid models, the gravitino mass scale is
  typically in the range of 
  $100$ GeV to  1 TeV, whereas in mirage mediation 
  a 100\,TeV gravitino is required to obtain a TeV-scale SUSY
  spectrum \cite{Lebedev,deflected}. 
}
\end{itemize}

\section{Gravitino dark matter in hybrid models}

As we emphasized in the introduction,
in the hybrid models that we are studying both the neutralino and
gravitino can be the LSP.
Indeed, Eq.(\ref{general}) shows that, depending on the value of
$\alpha$, 
the LSP can be either a neutralino (for small values of $\alpha$)
or a gravitino (when $\alpha$ increases).
Notice that the scalar soft breaking
terms are dominated by the flavor dependent gravity mediation in the
former case and flavor-blind gauge mediation in the latter.
We investigate here the possibility that the gravitino LSP is a viable 
dark matter.

In scenarios with gravitino dark matter the late decay of the NLSP
into the LSP produces electromagnetic and hadronic
showers. If the decay takes place after Big Bang
nucleosynthesis (BBN), the products of these showers may alter the
primordial
abundances of light elements \cite{gravitinobbn}.
Also, the late injection of electromagnetic energy may distort the
frequency dependence of the cosmic microwave background
spectrum from its observed blackbody
shape
\cite{hu,gravitinoTP1,pdg02}.

It has been shown
that hadronic BBN constraints
rule out the possibility of neutralino NLSP for gravitino masses above
$m_{3/2} \gtrsim 100$ MeV
\cite{gravitinoCMSSM,Feng:2004mt,Roszkowski:2004jd,Cyburt:2006uv,gravitinoD,Pradler:2006qh}.
However, if the
NLSP
is the lightest stau, $\widetilde \tau_1$,
we should
also take into account the effect of 
bound-states effects on the primordial $^6$Li
abundance. Indeed,
it has been shown \cite{reaction} that bound-state formation of
$\tilde \tau_1^-$ with $^4$He
can lead to an overproduction of $^6$Li via the catalyzed BBN (CBBN)
reaction
$^4$He $X^-$ $+$ D $\rightarrow$ $^6$Li + $X^-$
\cite{Kohri:2006cn,Kaplinghat:2006qr,Hamaguchi:2007mp,Jedamzik:2007qk,Li,bailly},
which has a serious impact on the regions with viable gravitino dark
matter \cite{Cyburt:2006uv,Dilution,Li,Kersten:2007ab,reference}.  
In fact, the observationally inferred upper limit on the primordial
$^6$Li abundance \cite{gravitinobbn} implies a stringent
upper bound on the stau NLSP lifetime
\begin{equation}
  \tau_{\widetilde \tau_1}\lsim 5\times10^{3}\ {\rm s}\ .
  \label{bbnlife}
\end{equation}
A similar bound can be extracted using the same arguments to avoid
overproduction of $^9$Be (see, e.g.,
\cite{Li,reaction}).
In the case of a stau
NLSP decays, the stau decays primarily
to the gravitino and a
$\tau$ lepton at tree level, via gravitational interactions with a
lifetime \cite{gravitinoCMSSM,Feng:2004mt}
\begin{equation}
  \tau_{\widetilde{\tau_1}}\simeq
  \Gamma^{-1}(\tilde \tau_1 \rightarrow \tilde G\tau)=
  6.1\times10^{6}\left(\frac{m_{\widetilde{G}}}
  {100\,\mathrm{GeV}}\right)^2
  \left(\frac{100\,\mathrm{GeV}}{m_{\widetilde{\tau}}}\right)^5
  \left(1-\frac{m_{\widetilde{G}}^2}
       {m_{\widetilde{\tau}}^2}\right)^{-4}\mathrm{s}\ .
       \label{staulifetime}
\end{equation}

The relic abundance of gravitinos receives contributions from two
different sources. First, there is a non-thermal production (NTP)
\cite{Feng:2003xh,Feng:2003uy,Feng:2004mt} of
gravitinos in the late decays of the NLSP.
Since each NLSP decays into one
gravitino, the non-thermal relic abundance of the latter is
related to that of the NLSP 
\cite{nonthermal,Steffen:2007sp}
\begin{equation}
  \Omega_{\tilde G}^{\mathrm{NTP}}h^2 =
  \frac{m_{3/2}}{m_{\mathrm{NLSP}}}\Omega_{\mathrm{NLSP}}^{\mathrm{TP}}h^2
  \sim
  0.02 \left( \frac{m_{3/2}}{100~\mathrm{GeV}} \right)
  \left( \frac{m_{\mathrm{NLSP}}}{1~\mathrm{TeV}}\right)\ .
  \label{NTP}
\end{equation}
Second, gravitinos are also thermally produced (TP)
through scatterings in the plasma, the resulting relic abundance being
proportional to the
reheating
temperature, $T_R$, of the Universe after inflation
\cite{Bolz:2000fu,Pradler:2006qh}
\begin{equation}
  \Omega_{\tilde G}^{\mathrm{TP}}h^2 \simeq 0.32 \left(\frac{100\,
    \mathrm{GeV}}{m_{3/2}} \right)
  \left(\frac{m_{\mathrm{\tilde g}}}{1\, \mathrm{TeV}} \right)^2
  \left(\frac{T_R}{10^7\, \mathrm{GeV}} \right)\ .
  \label{TP}
\end{equation}
The total relic density is the sum of both contributions
$  \Omega_{\tilde G}h^2
  =\Omega_{\tilde G}^{\mathrm{NTP}}h^2 +\Omega_{\tilde
    G}^{\mathrm{TP}}h^2$,
to which we will apply the constraint extracted from the WMAP data
\cite{wmap5yr}.

Thus, in order to check the viability of gravitino dark matter in
hybrid models, we have studied the parameter space, imposing the upper
bound of Eq.(\ref{bbnlife}) to the stau lifetime and the WMAP
constraint to
the relic abundance. We show that these stringent conditions can
be realized in a very particular region of the parameter space.

It should be stressed that since in these hybrid scenarios the
messenger scale is much larger than in the standard GMSB models, the
cosmological effect of  mediators is completely different.
Concerning the influence of heavy messengers on the reheating
temperature, as underlined by the authors of \cite{Dilution}, if the
post-inflationary reheating temperature is larger
than the mass of the lightest messenger, $M_{\tilde M1}$, the
"messenger number" which ensures the 
stability of the lightest messenger
should be violated, otherwise
the messenger population would overclose the universe if 
$M_{\tilde  M1} \gtrsim 30$ TeV.
The consequences of the decays of messengers after their freeze out
temperature, would be the dilution
of the dark matter gravitino component by the late time increasing of
the entropy. However, in hybrid models, GUT-scale messenger masses 
are well
above $T_R$ and its
population is naturally suppressed by the Boltzman factor in the
primordial thermal bath. Thus they should not have any late time
effects on the dark matter population.

\subsection{Results}

We have performed a scan in the parameter space of a
general hybrid model. For concreteness, we have fixed the number of
mediators to $N=6$, which leaves only two input parameters in the
equations describing the soft masses in Eq.\,(\ref{general}),
namely the gravitino mass and
the parameter $\alpha$.
Moreover, we also fixed the effects coming from non-standard
contributions, taking 
a typical value for $\delta = - 1.8$. As stressed in Section~2, a more detailed 
analysis taking into account a scan over $\delta$ could be useful in order to  
investigate the  little hierarchy problems in this
 kind of model, but this is beyond the scope of the paper.

We have calculated the low energy spectrum solving the renormalization
group equations with the code {\tt SPheno} \cite{Porod:2003um}, taking
into account the LEP constraints on the masses of
supersymmetric particles. We also included the current experimental
bounds on low energy observables, such as on the branching ratios of
rare decays
$b\to s\gamma$ and $B_S\to\mu^+\mu^-$.
In particular, we imposed
$2.85\times10^{-4}\le\,{\rm BR}(b\to s\gamma)\le 4.25\times10^{-4}$,
which is obtained
from the experimental world average
reported by the Heavy Flavor Averaging Group \cite{bsgHFAG07},
and the theoretical calculation in the Standard Model
\cite{bsg-misiak},
with errors combined in quadrature.
We have also taken into account the upper constraint on
the $(\bmumu)$ branching ratio obtained by CDF,
BR$(\bmumu)<5.8\times10^{-8}$ at $95\%$ c.l. \cite{bmumuCDF07}.
Given the current discrepancy between the experimental measurements of
the muon anomalous magnetic moment, $a_\mu\equiv(g-2)_\mu$,
from $e^+e^-$ or tau data,
we have not imposed any constraint on the resulting
supersymmetric correction, $\asusy$.
We nevertheless comment on the regions
which are favoured by the current $e^+e^-$ result \cite{g-2}
and the present evaluations of the Standard Model contributions
\cite{g-2_SM,newg2,kino}, which lead to
$\asusy=(27.6\,\pm\,8)\times10^{-10}$.

An important part of the $(\alpha,\,m_{3/2})$
plane of hybrid models was already studied in
Ref.\,\cite{Dudas:2008qf},
where regions with viable neutralino dark matter
were obtained with $1\lsim\alpha\lsim 8$ and moderate values
of the gravitino mass. Here we are interested in exploring a
complementary region of the parameter space,
with larger values of $\alpha$, where the gravitino is the LSP and
therefore a potential dark matter candidate.

As explained in the previous section, BBN constraints strongly
disfavour the regions with neutralino NLSP
\cite{Feng:2004mt,Roszkowski:2004jd,Cyburt:2006uv,gravitinoD,Pradler:2006qh}
and consequently we have excluded
these from the parameter space and studied only the regions 
with stau NLSP. In
these areas, the lifetime of the stau is calculated
and condition (\ref{bbnlife}) is used as an extra constraint.
Finally, the relic abundance of gravitinos is evaluated with the code
{\tt micrOMEGAs} \cite{micromegas} and
bounded using the WMAP result \cite{wmap5yr}.

\begin{figure}[!t]
  \epsfig{file=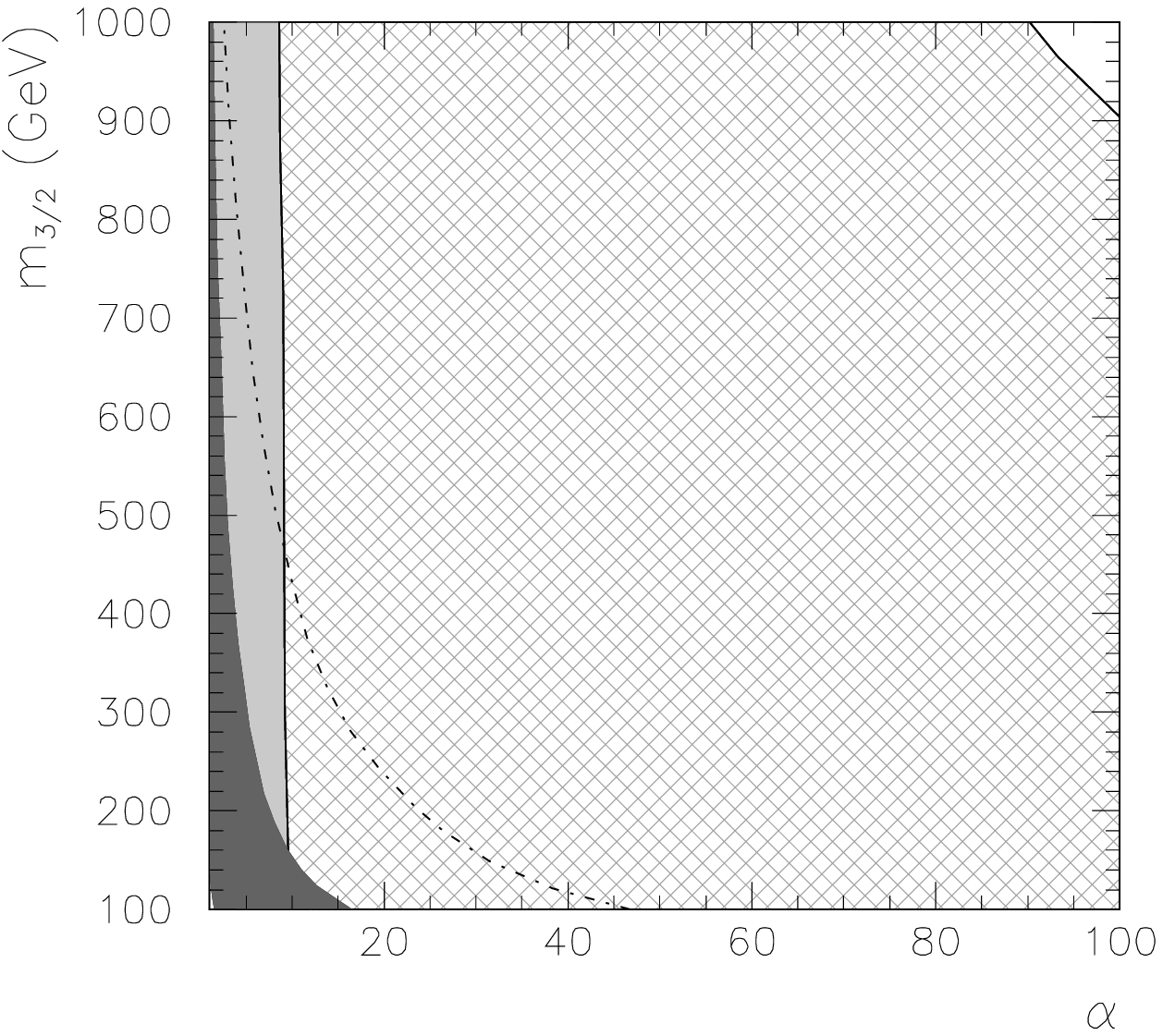,width=8cm}
  \epsfig{file=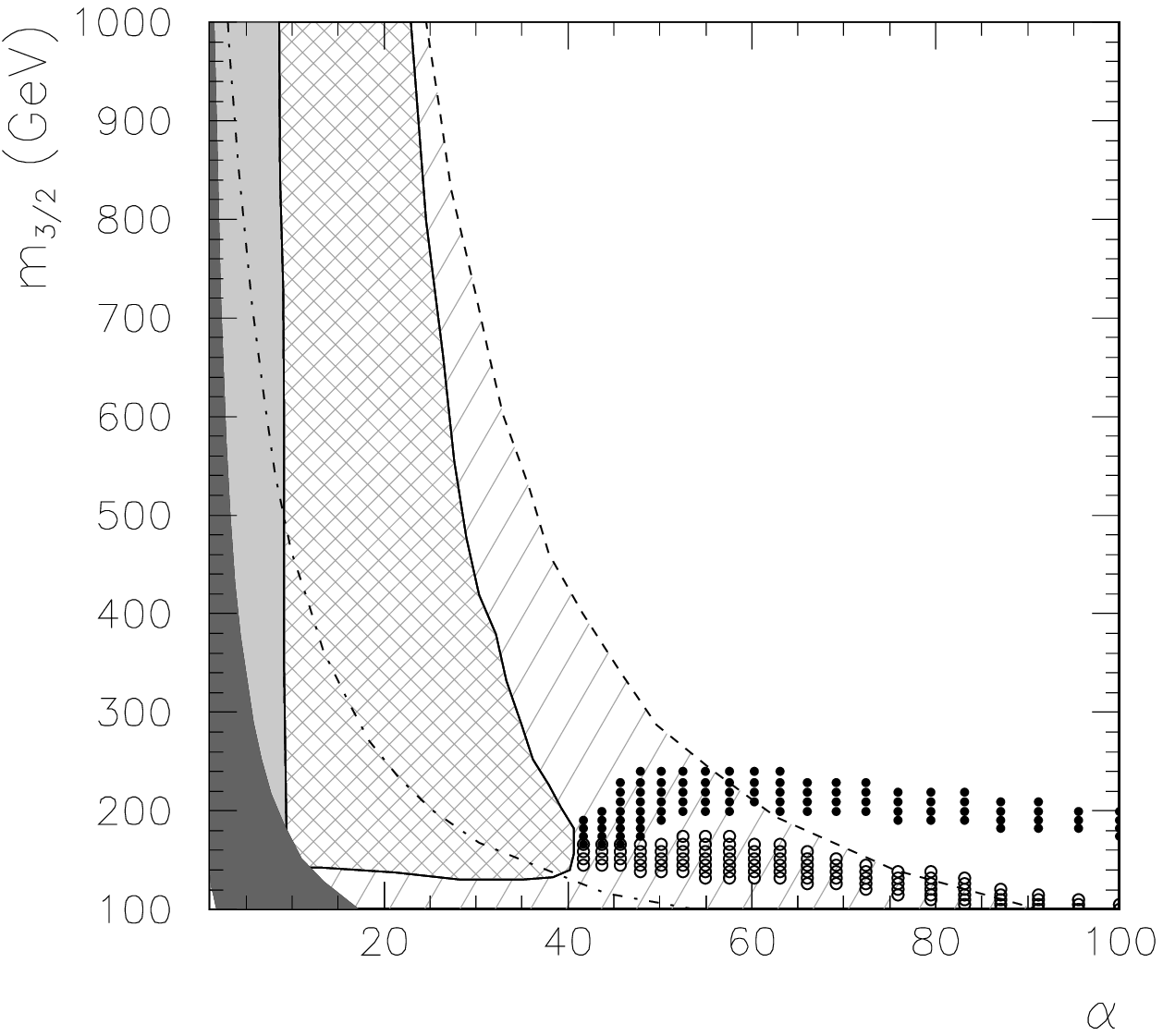,width=8cm}\\
  \epsfig{file=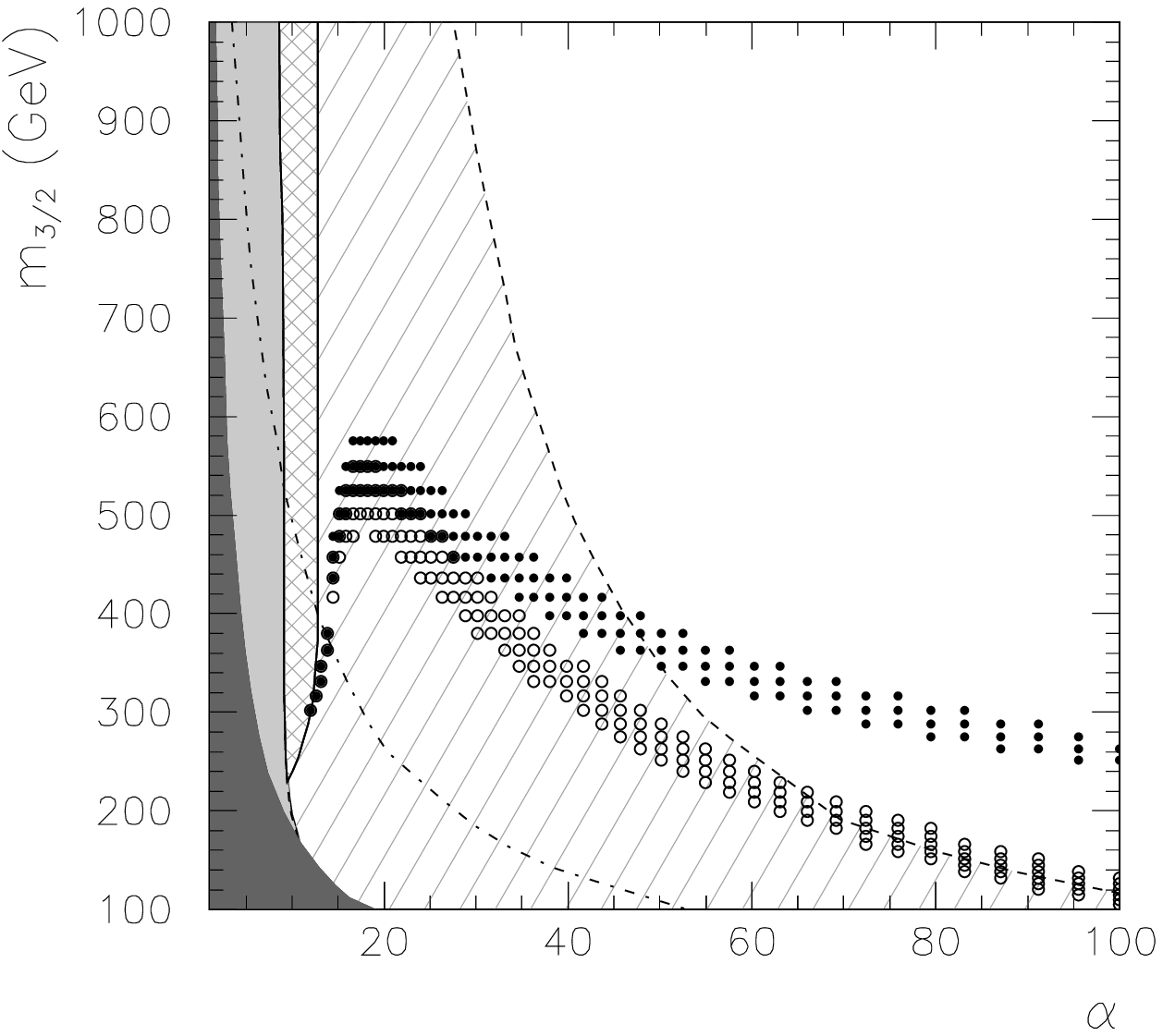,width=8cm}
  \epsfig{file=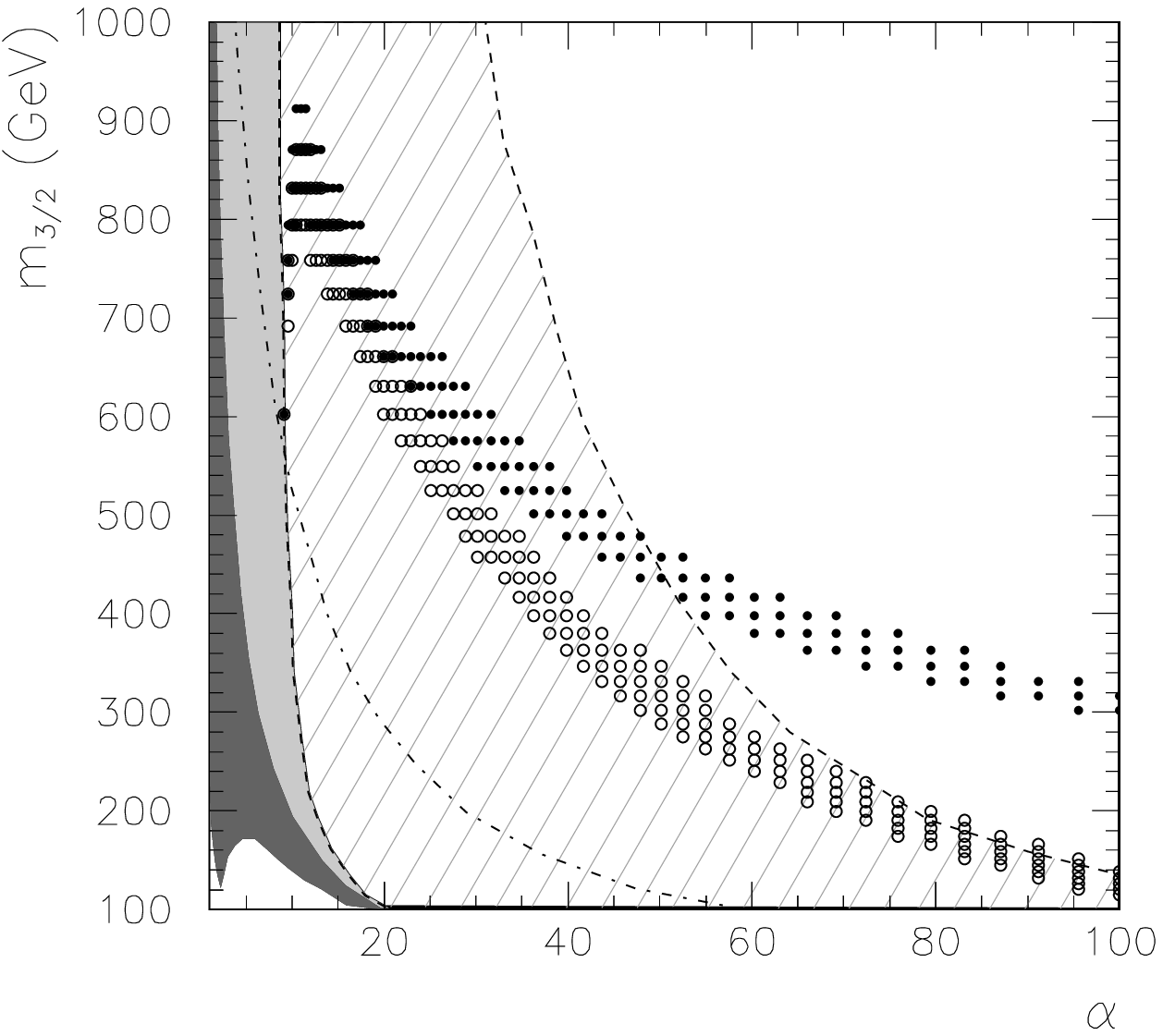,width=8cm}
  \captions{
    Gravitino mass as a function of $\alpha$ for $A=0$ and
    $\tan\beta=35,\,40,\,45$, and $50$ from left to right and top to
    bottom. The coefficient for non-standard GMSB contributions is
    always taken $\delta = - 1.8$. The line and colour code is
    explained in the text.}
  \label{malpha_0}
\end{figure}

\begin{figure}[!t]
  \epsfig{file=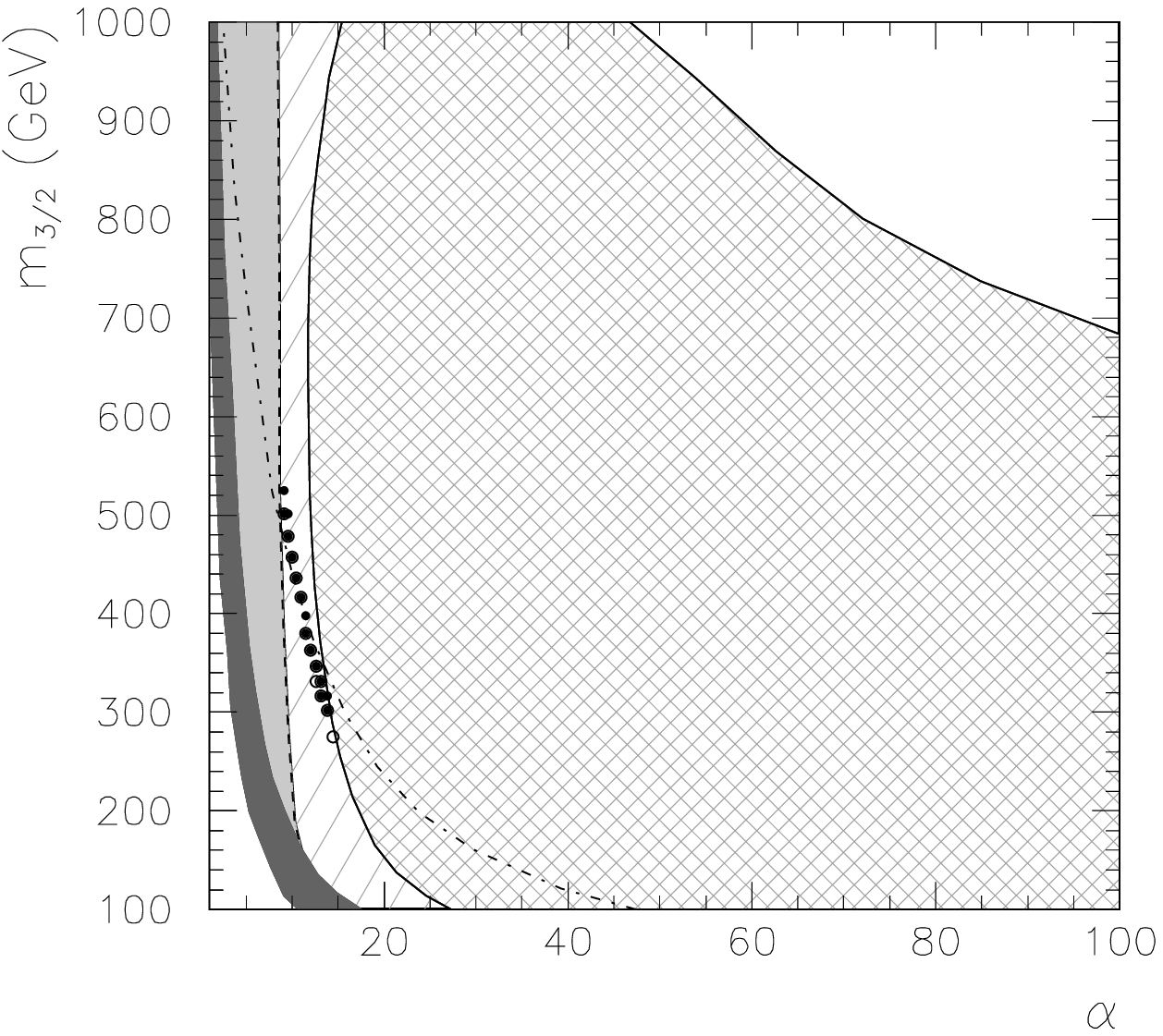,width=8cm}
  \epsfig{file=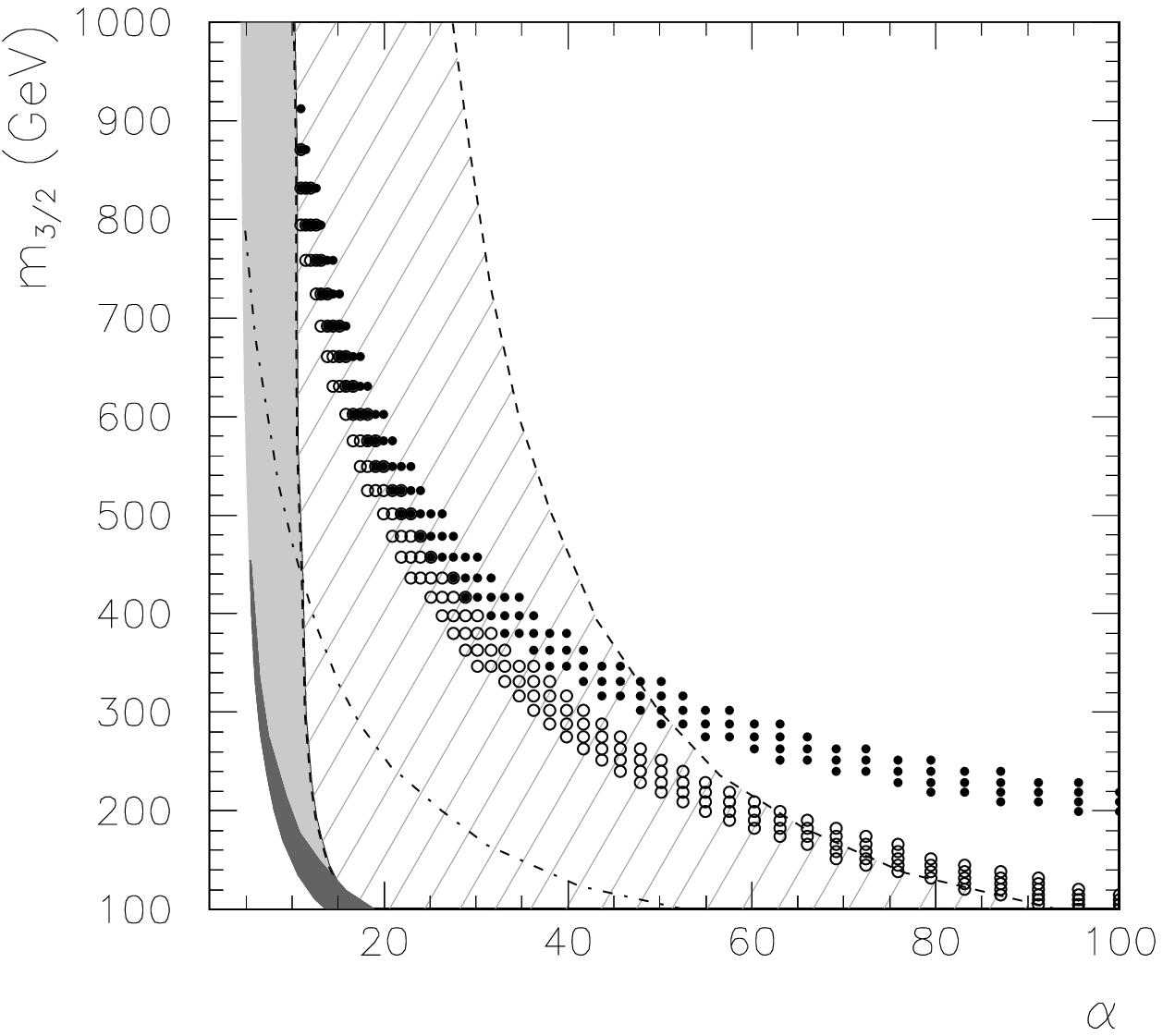,width=8cm}\\
  \epsfig{file=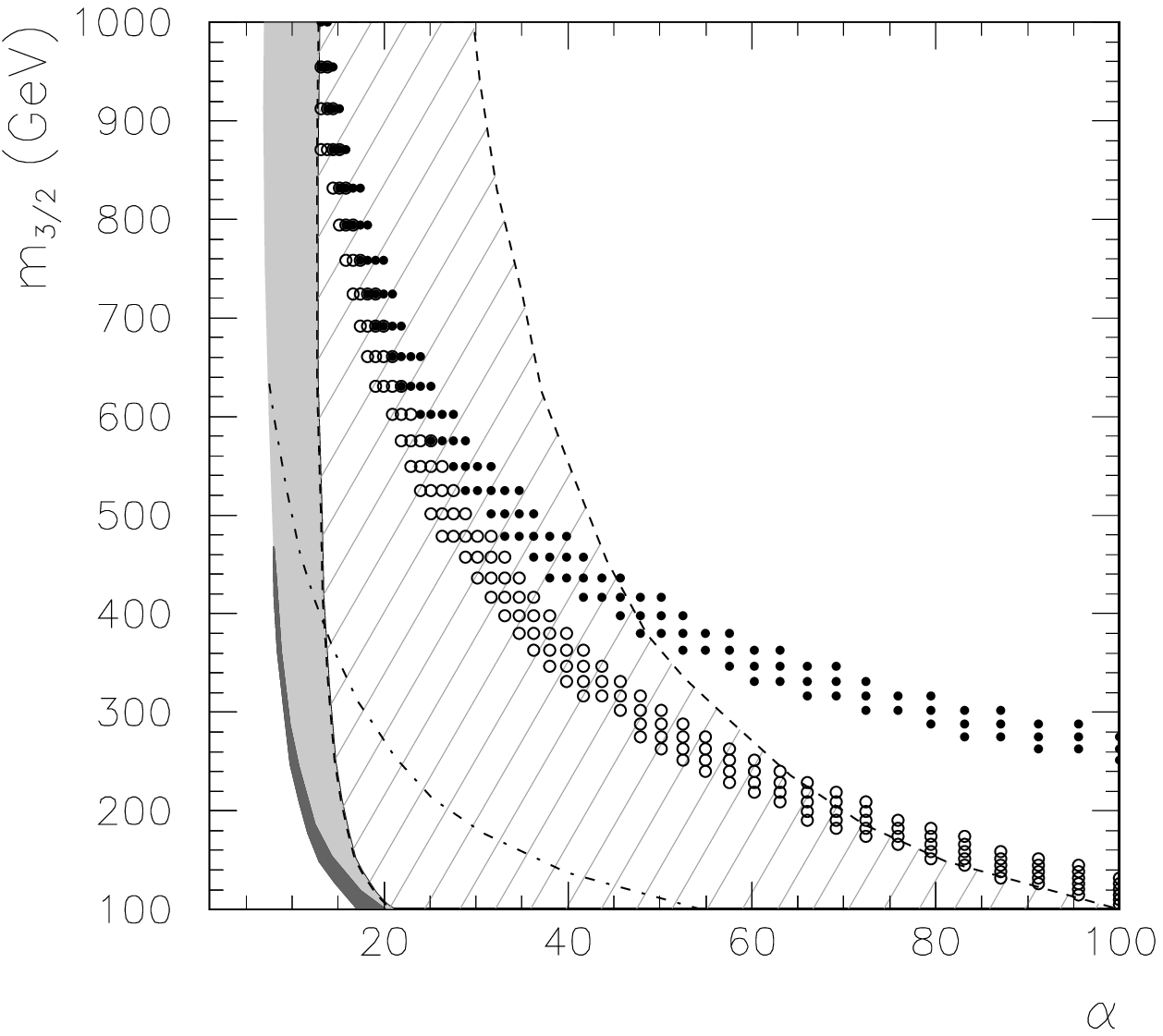,width=8cm}
  \epsfig{file=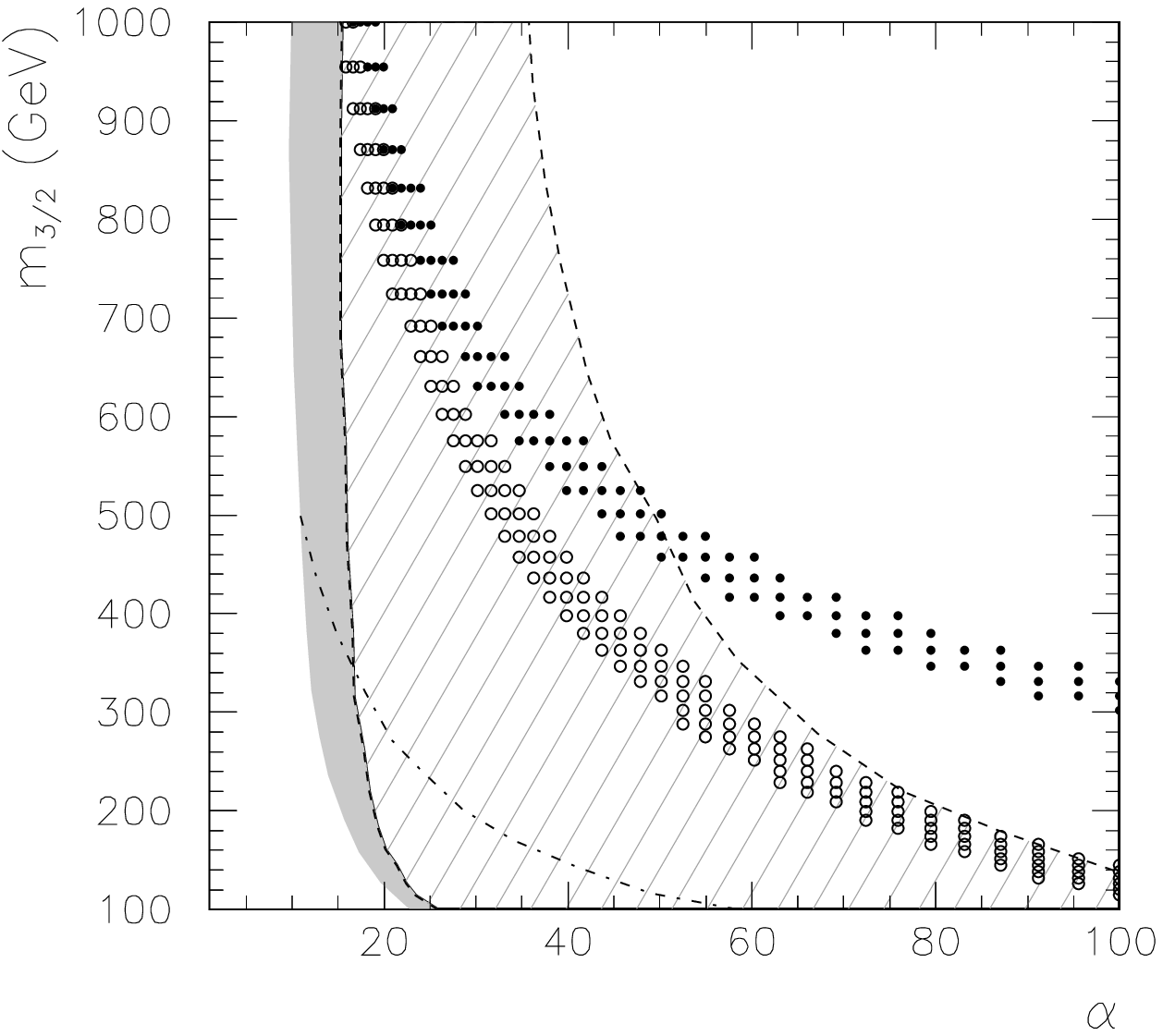,width=8cm}
  \captions{The same as in Fig.\,\ref{malpha_0} but for $A=-3\,m_{3/2}$.}
  \label{rmalpha_-3}
\end{figure}

As a first example, we fixed $A=0$ and performed a scan in the
$(\alpha,\,m_{3/2})$ plane for various choices of $\tan\beta$. The
results are displayed in Fig.\,\ref{malpha_0}.
As already shown in Ref.\,\cite{Dudas:2008qf},
the gravitino in hybrid models
becomes the LSP for $\alpha\gsim 10$, due to the increase
in both the gaugino and scalar mass parameters at the GUT scale
(see Eq.(\ref{general})). In the figure the light grey area
indicates the points in which the gravitino is not the LSP.
The region with gravitino LSP
corresponds to the area on the right of the almost vertical 
line\footnote{
  To the left of this line the stau (neutralino) is the LSP for
  light (heavy) gravitino masses.}
at $\alpha\sim 10$. Within that region, the gridded area corresponds to
points with neutralino NLSP and, as argued before, is excluded.
In the remaining regions of the
parameter space the stau is
the NLSP. The ruled area corresponds to the points in which the
condition in the stau lifetime given by Eq.(\ref{staulifetime}) 
is not fulfilled and we therefore
also consider it excluded by BBN constraints. Finally, in the
remaining white area (towards large values of $\alpha$) the stau
decays rapidly enough and BBN constraints are satisfied.

In the dark grey area at least one
experimental constraint is violated (LEP lower
bound on the Higgs or SUSY masses, and
branching ratios of rare decays BR($b\to
s\gamma$) and BR($B_S\to \mu^+\mu^-$)). Notice
that these are generically confined to the area with neutralino LSP,
since the spectrum is lighter.
Finally, to the right of the dot-dashed line, the supersymmetric
contribution to the muon anomalous magnetic moment would be too small
to account for the observed $e^+e^-$ data. Notice however that it
would not necessarily be in contradiction with tau data.
In Fig.\,\ref{rmalpha_-3} we represent the same example but for a
different soft trilinear term, $A=-3 m_{3/2}$.

A first thing to notice is that large values for
$\tan\beta$ are needed
in order to obtain regions of the parameter space in
which the stau is the NLSP instead of the neutralino. 
Indeed, 
for large $\tan\beta$ the bottom Yukawa increases, thereby
inducing a larger negative contribution to the running of the stau
mass parameters.
As we see in both
cases, $A=0$ and $A=-3\,m_{3/2}$, a value of
$\tan\beta \gsim 40$ is enough.
Also the areas with stau NLSP are wider for $A=-3\,m_{3/2}$
since the negative contribution to the running of the stau mass
parameters and the larger off-diagonal elements in the mass matrix
(which also increase with $\tan\beta$)  
imply lighter staus.

As discussed previously, the most stringent
constraint stems from $^6$Li and $^9$Be overproduction. In order for
the stau lifetime to be short enough, the stau has to be sufficiently
heavy (while still being lighter than the lightest neutralino).
For a given gravitino mass $m_{3/2}$, this implies a lower bound on
$\alpha$.
This can be qualitatively understood from Eq.(\ref{staulifetime}) and
Eq.(\ref{general}) as follows.
An increase in $\alpha$ for fixed $m_{3/2}$ implies an increase in the
stau mass and consequently a decrease in its lifetime.
For small values of $\beta$ and fixed $N$ we can
approximate $\tau_{\tilde \tau_1}\propto \alpha^{-\frac{5}{2}}
m_{3/2}^{-3}$, which is in agreement with the slope of the dashed line
in 
Fig.\,\ref{malpha_0}.

Regarding the resulting relic abundance, we display two examples with
different values for the reheating temperature.
Black dots represent the results with $T_R=10^6$~GeV and empty circles
correspond to $T_R=10^8$~GeV.
In fact, for $T_R=10^6$~GeV the thermal contribution is
very small and these points can be understood as coming from purely
non-thermal production.
On the one hand,
from Eq.(\ref{TP}) one can qualitatively infer that for a
fixed number of messengers the thermal production can be approximated
as  $\Omega_{\tilde G}^{\mathrm{TP}}h^2 \propto m_{3/2}
\alpha^2$. This is consistent with the slope of the region with
$T_R = 10^8$ GeV where the relic abundance of
gravitino is mainly thermal.
On the other hand, using the same qualitative arguments,
the non-thermal
contribution to the gravitino relic abundance should behave as
$\Omega_{\tilde G}^{\mathrm{NTP}}h^2 \propto m_{3/2}^2 \alpha$.

\begin{figure}[!t]
  \begin{center}
    \epsfig{file=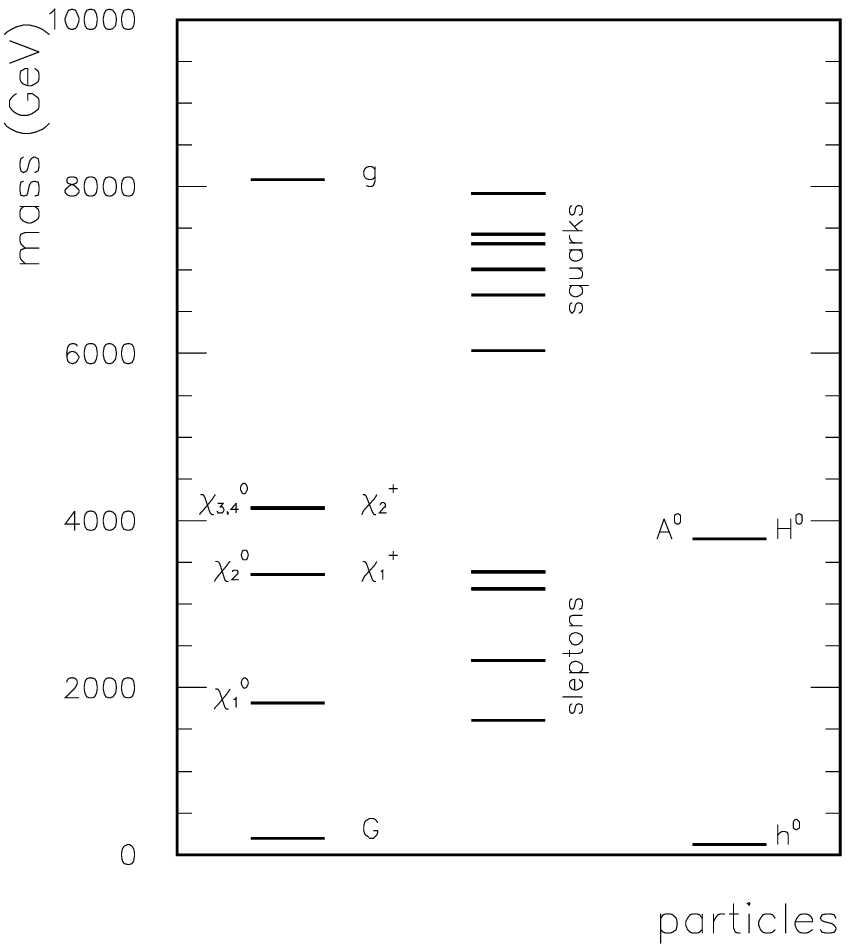,width=8cm}
  \end{center}
  \captions{Supersymmetric spectrum for a representative example in the
  parameter space with $A=0$,
  $\tan\beta=45$, $\alpha=80$ and $m_{3/2}=200$~GeV
  for which viable gravitino dark matter is obtained.}
  \label{spectrum}
\end{figure}

As already pointed out 
\cite{Feng:2003xh,Feng:2003uy,gravitinoCMSSM,Feng:2004mt,reference} 
within the context of the
CMSSM, in order to fulfil the BBN constraints the mass of the light
stau needs to be $m_{\tilde \tau}\gsim2$~TeV. 
The regions with viable gravitino dark matter
that lie between the two limiting values of $T_R$, correspond in our
case to
$m_{\widetilde \tau_1} \propto m_{3/2}\, \alpha \gsim1.6$~TeV.
An example of the characteristic spectrum that would be expected in
these models is shown in Fig.\,\ref{spectrum} for $A=0$ and
$\tan\beta=45$ with $\alpha=80$ and $m_{3/2}=200$~GeV.
As expected, the spectrum is very heavy, only the lightest stau
and the lightest neutralino have masses below $2$~TeV and the whole
squark sector above $6$~TeV.

To sum up, regions with viable gravitino dark matter can be found in
this general class of hybrid models.
They correspond to areas with a large value of $\alpha$, of order
$45$ (typically this implies $\langle X \rangle \lesssim 10^{16}$), with
gravitinos in the mass range of
several hundred GeV and
$T_R\lsim10^8$~GeV.
The rest of the supersymmetric spectrum is rather heavy,
with staus in the mass-range of 2~TeV and with
$\tan\beta\gsim 40$.

Therefore the allowed region in the parameter space corresponds to the
opposite of the range considered in
\cite{Dudas:2008qf}, where a very small value of the Fayet-Iliopoulos
term is required and GMSB begins to be dominant. In particular, the
construction discussed there seems to slightly disfavor these points
with gravitino dark matter, even if still possible in the framework of
such models.

Another interesting point would be to see some collider 
signatures of such
class of models.
Indeed, the stau lifetime in Eq.(\ref{staulifetime}) depends on the
messengers mass through $F_X$,
contrary to secluded sector breaking. Thus, any information on the
stau lifetime would
give information on the
messenger mass scale. For instance, 100\,TeV messengers imply a short stau
lifetime,
whereas $10^{16}$ GeV messengers mass lead to a stau lifetime of
seconds.
Variations of orders of magnitude
in the messenger masses directly induce differences of 
orders of magnitude in the
stau lifetime \cite{TDRATLAS}.

\section{Conclusions}

In this work we have studied the phenomenology of a generic class of 
string motivated scenarios in which gravity
mediation naturally competes with gauge mediation as the origin of
supergravity-breaking. An interesting feature of these constructions
is that the messenger masses are of order of the GUT scale, contrary
to standard GMSB models. 
In these scenarios the neutralino is
typically the lightest supersymmetric particle when gauge and gravity
contributions are of the same order.
However, when gauge-mediation becomes dominant, the gravitino easily
becomes the LSP and therefore a potential dark matter candidate. 
We have shown that even without secluded breaking sector, heavy
messengers induce indirectly
a GeV/TeV gravitino mass if the contribution to the cosmological
constant comes from the spurion field.

We
have then investigated the viability of the gravitino as dark matter, 
calculating its relic
abundance and imposing the WMAP result. 
Furthermore, we have taken into account existing bounds 
from Big Bang nucleosynthesis. The latter play a leading role in
constraining the parameter space. 
Regions with viable gravitino dark matter
can be obtained when the SUSY breaking mechanism is mostly dominated
by gauge mediation ($\alpha\gsim 45$) and with
$\tan\beta \gsim 40$.
The resulting spectrum is relatively heavy, with squark masses 
larger or of order of $6$~TeV and slepton masses above $2$~TeV.

\vspace*{1cm}
\noindent{\bf Acknowledgements}

We thank S.~Abel and E.~Dudas for useful discussions and continuous 
support. 
D.G.C. was supported by the program
``Juan de la Cierva'' of the Spanish MEC
and also in part by the Spanish DGI of the
MEC under Proyecto Nacional FPA2006-01105,
by the EU
network
MRTN-CT-2006-035863 and the project HEPHACOS P-ESP-00346
of the Comunidad de Madrid.
The work of A.R. was  supported
by the European Commission Marie Curie Intra-European
Fellowships under the contract N 041443.
We also thank the ENTApP Network of the
ILIAS project RII3-CT-2004-506222.
The work is also sponsored by the hepTOOLS Research Training Network
MRTN-CT-2006-035505.

\end{document}